\documentclass[aps,twocolumn,pra,superscriptaddress,showpacs,tightenlines]{revtex4}
\usepackage{amssymb}
\usepackage{amsmath}
\usepackage{graphicx}
\usepackage{epsfig}
\usepackage{subfigure}
\usepackage{amsfonts}

\begin{document}

%(when sumit to arxiv, remove \begin{CJK*}{GBK}{song})
\title{Quantum decoherence in a hybrid atom-optical system of a one-dimensional coupled-resonator waveguide and an atom}
\author{Jing Lu}%(when sumit to arxiv, remove (¬¾º))
\affiliation{Key Laboratory of Low-Dimensional Quantum Structures
and Quantum Control of Ministry of Education, and Department of
Physics, Hunan Normal University, Changsha 410081, China}
\author{Lan Zhou}%(when sumit to arxiv, remove (ÖÜÀ¼))
\thanks{Corresponding author}
\email{zzhoulan@gmail.com} \affiliation{Key Laboratory of
Low-Dimensional Quantum Structures and Quantum Control of Ministry
of Education, and Department of Physics, Hunan Normal University,
Changsha 410081, China}
\author{H. C. Fu}%(when sumit to arxiv, remove (¸¶ºé³À))
\affiliation{School of Physics Science and Technology, Shenzhen
University, Shenzhen 518060, P.R.China}
\author{Le-Man Kuang}%(when sumit to arxiv, remove (¿ïÀÖÂú))
\affiliation{Key Laboratory of Low-Dimensional Quantum Structures
and Quantum Control of Ministry of Education, and Department of
Physics, Hunan Normal University, Changsha 410081, China}

\begin{abstract}
Decoherence for a one-dimensional coupled-resonator waveguide with a two-level
 system inside one of resonators, induced by their interaction with
 corresponding environments, is investigated. Each environment is modeled as
 a continuum of harmonic oscillators.
 By finding the eigenstates of the hybrid system, which is the dressed state of the hybrid system,
 we calculate the lifetime of one excitation, which characterizes the existence of quantum coherence
 in such hybrid system and the basic quantum nature.
\end{abstract}

\maketitle \narrowtext %(when sumit to arxiv, remove \end{CJK*}%
\narrowtext

\section{\label{Sec:1}Introduction}

Any discrete state coupled to continua of states is subject to
decay. The intrinsic dynamics of such quantum system is
irreversible, i.e. once a system is initially in the discrete state,
it never returns to the initial state spontaneously. Such kind of
phenomenon is also described by the so-called resonant tunneling
process. There are many examples of such processes in different
branches of physics. A typical example is an excited two-level
system interacting with the modes of electromagnetic fields.

Spontaneous emission is normally regarded as a loss and a
decoherence mechanism. Recent works show that spontaneous emission
of the excited two-level system can be exploited to influence the
coherent transport properties of single photon in a one-dimensional
(1D) waveguide due to the interference between the spontaneous
emission from two-level systems and the propagating modes in the 1D
continuum~\cite{PRL95(05)213001}. For a discrete system, the
simplest model which possesses the resonant tunneling process is the
so-called Anderson-Fano-Lee model~\cite
{Fano124(61),Anderson124(61),Lee95(54),PRA74(06)032102}, in which
the continuum is formed by a linear chain of sites with the
nearest-neighbor interaction and the rate of emission is modified
due to the change of the density of state.

With the
development of microfabrication technology, the platforms, such as
defect resonators in photonic crystals~\cite
{JAP75(94)4753,PRB54(96)007837,OL24(99)711,APL84(04)161} and coupled
superconducting transmission line resonators~\cite
{YNPRB68(03),n431(04)162,n431(04)159,PRA77(08)013831}, are promising
candidates for realizing a waveguide (or an
array)~\cite{n424(03)817}, under the tight-binding approximation.
Since a 1D coupled-resonator waveguide
(CRW) possesses a band-gap spectrum and can transmit a wave packet
of light, a single-photon quantum
switch~\cite{PRL101(08)100501,PRA78(08)053806}, made of a
controllable two-level or three-level system, has been studied using
a discrete-coordinate approach.
It is then found that the
transmission of a single-photon in a 1D CRW can be switched on and
off by modulating the energy-level spacing of the two-level system
(TLS) with a high-frequency signal~\cite{PRA80(09)062109}.

The system with TLSs inside a 1D CRW~\cite%
{PRL101(08)100501,PRA78(08)053806,PRA80(09)062109,PRA78(08)063827},
is investigated under the condition of ideal resonators and ideal
couplings of the resonator modes to the respective atomic
transitions. Such a system is closed, and the existence of
superpositions prescribed by quantum mechanics is valid. All the
results in Refs.~\cite{PRL101(08)100501,PRA78(08)053806} are
obtained by seeking the stationary states of the system with a TLS
inside 1D CRW. However a realistic quantum system can rarely be
isolated from its surrounding completely, rather it is usually
coupled to the external environment (also called ``heat bath'' )
with a large number of degrees of freedom. There are two main loss
processes which are serious obstacles against the preservation of
quantum coherence over long period of time: spontaneous emission
from the excited state to the ground state due to its interaction
with the modes outside the resonator, and leaking out of photons of
the resonator mode. Obviously, when the external environment is
taken into account, the decoherence of every resonator and the TLS
would result in the incoherent or dissipative propagation of the
incident photon.

It is pointed in Refs.~\cite
{PRL101(08)100501,PRA78(08)053806,PRA80(09)062109} that the
decoherence or dissipation can be divided into two categories: one
influences the free propagation of the single photon, and another
influences the scattering process which broadens the line width.

Therefore in the present work, we shall investigate the influence of
environment to the decoherence of 1D coupled resonator with a
two-level system inside. We mainly focus on two issues, one is the
lifetime of each eigenstates characterizing the time-scale of
quantum coherence in a 1D CRW with a TLS inside, and another is the
the rate of decay through introducing the leakage rate in each
resonator and the decay rate of the TLS, which influences the free
propagation of the single photon. The reflection spectrum is also
obtained and it is found that the dissipation lowers the peak of the
resonance and broadens the line width. However, the total reflection
can still be achieved when the leakage rate in each resonator is
equal to the decay rate of the TLS.

This paper is organized as follows. In Sec.~\ref{Sec:2}, we
introduce a microscopic model where both the resonators and the TLS
are coupled to its own surrounding through an exchange interaction.
In Sec.~\ref{Sec:3}, we briefly review the stationary states of a
CRW with a TLS inside in Ref.~\cite{PRL101(08)100501}. In
Sec.~\ref{Sec:4}, we derive the characteristic time for a
single-photon staying in the system with a TLS embedded in 1D CRW.
In Sec.~\ref{Sec:5}, we investigate the impact of dissipation on
scattering amplitude. Conclusions are made in Sec. \ref{Sec:6}.

\section{\label{Sec:2} Model}

\begin{figure}[tbp]
\includegraphics[bb=29 542 389 770, width=8.5 cm,clip]{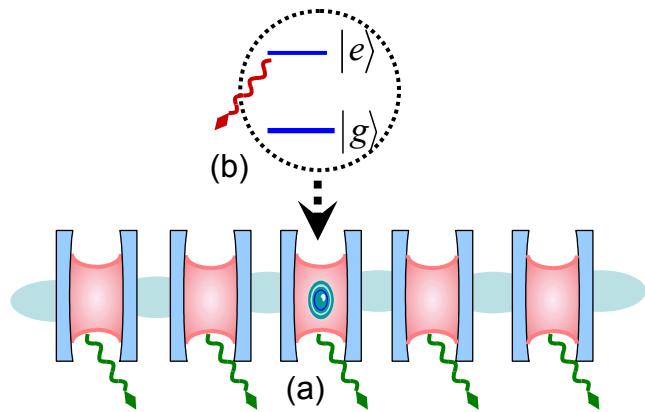}
\caption{(Color online) Schematic illustration of the model, where a two
level system (a) is inside a 1D coupled resonator waveguide (b). The two
level system and each of resonators are coupled to its corresponding
environment, which is modeled as an infinite set of harmonic oscillators.}
\label{fig2:1}
\end{figure}

Figure.\ref{fig2:1} shows a schematic diagram of what we consider in
this paper. The total system includes the system S and the
environment B. The environment B doesn't explicitly show in
Fig.\ref{fig2:1}. The system S refers to a 1D CRW with a TLS inside
one of the resonators as discussed in Ref.\cite {PRL101(08)100501}
and the environment B refers to all the subsystems interacting with
each resonator in the 1D CRW and the TLS. In Fig.\ref{fig2:1} the
wavy line with an arrow indicates that a bath interacts with this
subsystem.

Each resonator in the CRW is modeled as a harmonic oscillator mode
with frequency $\omega _{c}$. Due to the overlap of the spatial
profile of the resonator modes, photons can hop between neighboring
resonators. Introducing the creation and annihilation operators
$a_{j}^{\dagger }$ and $a_{j}$ for $j$th resonator, the Hamiltonian
for the CRW is given by
\begin{equation}
H_{\mathrm{C}}=\sum_{j}\omega _{c}a_{j}^{\dagger }a_{j}-\xi \sum_{j}\left(
a_{j}^{\dagger }a_{j+1}+\mathrm{H.c.}\right) \,,  \label{2a1}
\end{equation}%
where the intercavity coupling constant $\xi $ is the same for all
neighboring cavity-cavity interactions. A TLS with transition
frequency $\Omega $ is located in the 0th resonator. The
atom-resonator interaction is described by the Jaynes-Cummings model
\begin{equation}
H_{\mathrm{I}}=\Omega \left\vert e\right\rangle \left\langle e\right\vert
+J\left( \left\vert e\right\rangle \left\langle g\right\vert a_{0}+\mathrm{%
H.c.}\right) ,  \label{2a2}
\end{equation}%
under the dipole and rotating wave approximations. Here $\left\vert
e\right\rangle $ and $\left\vert g\right\rangle $ are the excited
state and ground state of the TLS, respectively. The dynamics of the
system S is governed by the Hamiltonian
\begin{equation}
H_{\mathrm{S}}=H_{\mathrm{C}}+H_{\mathrm{I}}.  \label{2a3}
\end{equation}%
By employing the Fourier transformation
\begin{equation}
a_{j}=\frac{1}{\sqrt{N}}\sum_{k}e^{ikj}a_{k},  \label{2a4}
\end{equation}%
Hamiltonian $H_{\mathrm{S}}$ is transformed into a k-space
representation
\begin{equation}
H_{\mathrm{S}}=\sum_{k}\Omega _{k}a_{k}^{\dagger }a_{k}+\Omega
\left\vert e\right\rangle \left\langle e\right\vert
+\frac{J}{\sqrt{N}}\sum_{k}\left( \left\vert e\right\rangle
\left\langle g\right\vert a_{k}+\mathrm{H.c.} \right)  \label{2a5}
\end{equation}%
where the lattice constant is assumed to be unity and $\Omega
_{k}=\omega _{c}-2\xi \cos k$ is the well-known Bloch dispersion
relation. Here, the Hamiltonian $H_{\mathrm{C}}$ given by a diagonal
matrix describes the extended states of the continuum. The third
term in Eq.(\ref{2a5}) is responsible for the interaction between
the TLS and the continuum. The Hamiltonian $H_{\mathrm{S}}$
describes that a quasiexcitation is created or annihilated in the
$k$-th mode, and the TLS transits from its excited state to the
ground state or vice versa. Obviously, it is easy for the TLS to
transit to its ground state, and very difficult to go back to its
excited state. Therefore, the CRW with a TLS inside is a typical
system of quantum dissipation phenomenon.

The environment B is modeled as a set of infinite number of harmonic
oscillators. In this paper, we assume each resonator of the CRW is
coupled to an individual bath of harmonic oscillators. The dynamic
of the $j$th resonator and its corresponding environment is governed
by the Hamiltonian
\begin{equation}
H_{\mathrm{ER}}^{[j]}=\sum_{q}\omega _{q[j]}b_{q[j]}^{\dag
}b_{q[j]}+\sum_{q}\left( g_{q[j]}a_{j}^{\dag }b_{q[j]}+\mathrm{H.c.}\right) ,
\label{2a6}
\end{equation}%
where $b_{q[j]}$ and $b_{q[j]}^{\dag }$ are the
annihilation and creation operators for the $q$th bath oscillator
attached to the $j$th resonator of the CRW, $\omega _{q[j]}$ is its
frequency, and $g_{q[j]}$ the coupling strength. The total
Hamiltonian describing the interaction between the CRW and
environments reads
\begin{equation}
H_{\mathrm{EC}}=\sum_{j}H_{\mathrm{ER}}^{[j]}.  \label{2a7}
\end{equation}%
We also introduce the exchange interaction between the TLS and its
environment, which is described by the Hamiltonian
\begin{equation}
H_{\mathrm{EA}}=\sum_{q}\nu _{q}d_{q}^{\dag }d_{q}+\sum_{n}\beta _{n}\left(
d_{n}^{\dag }\left\vert g\right\rangle \left\langle e\right\vert +\left\vert
e\right\rangle \left\langle g\right\vert d_{n}\right) ,  \label{2a8}
\end{equation}%
where $d_{n}$ and $d_{n}^{\dag }$ are the annihilation and creation
operators corresponding to the $q$th mode with frequency $\nu _{q}$.
The Hamiltonian of the total system S+B reads
\begin{equation}
H_{\mathrm{SB}}=H_{\mathrm{S}}+H_{\mathrm{EC}}+H_{\mathrm{EA}}.  \label{2a9}
\end{equation}%
Although there exists energy exchange between the system S and the
environment B, the total number of excitations is preserved.

\section{\label{Sec:3}Stationary states of a CRW with a TLS inside}

In the one-excitation subspace, the stationary states of system S
are either the localized states around the location of the TLS or a
superposition of extended propagating Bloch waves incident reflected
and transmitted by the TLS embedded in the CRW. Assume a photon is
coming from the left of the TLS with energy
\begin{equation}
\Omega _{k}=\omega _{C}-2\xi \cos k.  \label{3a1}
\end{equation}%
The stationary state of the system is then
\begin{equation}
\left\vert \Omega _{k}\right\rangle =\sum_{j}u_{k}\left( j\right)
a_{j}^{\dag }\left\vert 0g\right\rangle +u_{ek}\left\vert 0e\right\rangle ,
\label{3a2}
\end{equation}%
where $\left\vert 0\right\rangle $ is the vacuum state of the cavity
field, $u_{ek}$ is the probability amplitude with the TLS in the
excited state and no photon, and $u_{k}\left( j\right) $ is the
probability amplitude for the TLS to move to the lower state,
emitting a photon into a mode of the $j$th resonator. From the
eigenvalue equation $H_{\mathrm{S} }\left\vert \Omega
_{k}\right\rangle =\Omega _{k}\left\vert \Omega _{k}\right\rangle $,
together with the bosonic commutation relations, we derive a system
of coupled equations among the amplitudes
\begin{subequations}
\label{3a3}
\begin{eqnarray}
\left( V_{k}+JG _{k}\delta _{j0}\right) u_{k}\left( j\right)  &=&\xi
\left[ u_{k}(j+1)+u_{k}\left( j-1\right) \right] , \\
u_{ek} &=&G_{k}u_{k}\left( 0\right) ,
\end{eqnarray}%
with the Green function $G _{k}=G _{k}(\Omega )=J/\left( \Omega
_{k}-\Omega \right) $ and $V_{k}=\omega _{C}-\Omega _{k}$.
Equation~(\ref{3a2}) has the solution in terms of incoming and
outgoing waves with amplitudes
\end{subequations}
\begin{equation}
u_{k}\left( j\right) =\left\{
\begin{array}{ll}
e^{ikj}+r_{k}e^{-ikj}\text{, } & j<0 \\
s_{k}e^{ikj}\text{, \ \ \ \ \ \ }& j>0%
\end{array}%
\right.   \label{3a4}
\end{equation}%
The reflection and transmission amplitudes yield
\begin{subequations}
\label{3a5}
\begin{align}
r_{k}& =s_{k}-1, \\
s_{k}& =\frac{2i\xi \left( \Omega _{k}-\Omega \right) \sin k}{2i\xi \left(
\Omega _{k}-\Omega \right) \sin k-J^{2}}.
\end{align}%
Eq.(\ref{3a3}) implies that the transmission coefficient is sensitive to transition frequency
of the TLS, as the forward and backward propagating modes within the CRW are coupled via the TLS.

The periodicity of the CRW in $r$-space gives arise to a continuum
with Bloch waves grouped in energy bands, which is broken due to the
coupling between the TLS and the CRW. Local modes are produced, and
their corresponding eigenenergy is outside the energy bands. We
denote the eigenvalue of the bound state as $\Omega _{\kappa }$,
i.e. replace the index $k$ with $\kappa $, so does the equation
(\ref{3a2}). Noticing that the probability of of bound state at
infinity in coordinate space is zero, i.e. bound state is spatially
localized, we can assume that the bound states of even parity have
the following amplitudes
\end{subequations}
\begin{equation}
u_{\kappa }\left( j\right) =\left\{
\begin{array}{c}
Ce^{\left( in\pi -\kappa \right) j}\text{, }j>0 \\
Ce^{\left( in\pi +\kappa \right) j}\text{, }j<0%
\end{array}%
\right.  \label{3a6}
\end{equation}%
where the normalization constant $C$ is
\begin{equation}
C=\left[ \tanh k+\frac{J^{2}}{\left( \Omega _{\kappa }-\Omega \right) ^{2}}%
\right] ^{-1/2}.  \label{3a7}
\end{equation}%
The bound state energy lies either below or above the continuum with the
magnitude
\begin{equation}
\Omega _{\kappa }=\omega _{C}-2\xi e^{in\pi }\cosh \kappa ,  \label{3a8}
\end{equation}%
where the value of $\kappa $ is determined by the following condition%
\begin{equation}
J^{2}=2\xi e^{in\pi }\left( \Omega -\Omega _{\kappa }\right) \sinh \kappa .
\label{3a9}
\end{equation}%
The bound states exist when Eq.(\ref{3a9}) has solution.

\section{\label{Sec:4}Single-particle spontaneous-emission}

Quantum states inevitably decay with time into a probabilistic
mixture of classical states due to their interaction with the
environment, which comprises much larger systems or ensemble of
states. In this section we derive the relaxation time for single
combined photonic-atomic excitation in the system S by incorporating
the effects of interaction with the environment B. Notice that the
number of excitations is conserved in the total system (S+B).
Therefore, in one-excitation subspace, two situations will occur in
the system S, one is that the single-excitation is in state $
\left\vert \Omega _{k}\right\rangle $ or $\left\vert \Omega _{\kappa
}\right\rangle $; and another is that no excitation is found in the
system S, i.e. the TLS is in the ground state and no photon is in
the resonator. For the first case, there is no excitation in the
environment B, and we use the notation $\left\vert \Omega
_{k}00\right\rangle $ to describe the state of the total system,
where the first $0$ in the ket indicates that the radiation modes
coupled to the CRW are in the vacuum state, and the second $0$ in
the $\left\vert \Omega _{k}00\right\rangle $ denotes the vacuum
state of the harmonic oscillators attached to the TLS. When the
system S is in $\left\vert 0g\right\rangle $ state, the environment
B contains the single-excitation. Here, $d_{n}^{\dag }\left\vert
G\right\rangle $ ($\left\vert G\right\rangle \equiv \left\vert
0g00\right\rangle $) represents that the single-particle has moved
to the $n$th bath oscillator coupled to the TLS; and $b_{q[j]}^{\dag
}\left\vert G\right\rangle $ describes that the excitation in S is
in the $q$th mode of environment B attached to the $j$th resonator
of the CRW. Obviously, states $\left\{ \left\vert \Omega
_{k}00\right\rangle ,d_{q}^{\dag }\left\vert G\right\rangle
,b_{q[j]}^{\dag }\left\vert G\right\rangle \right\} $ provide a
complete basis and thus we can expand the wave function of the S+B
at arbitrary time $t$ in terms of this basis as
\begin{eqnarray}
\left\vert \Psi \left( t\right) \right\rangle &=&\sum_{k}U_{k}\left(
t\right) \left\vert \Omega _{k}00\right\rangle +\sum_{n}C_{n}\left( t\right)
d_{n}^{\dag }\left\vert G\right\rangle  \label{4a1} \\
&&+\sum_{qj}B_{q[j]}\left( t\right) b_{q[j]}^{\dag }\left\vert
G\right\rangle \notag,
\end{eqnarray}%
where $U_{k}$, $B_{q[j]}$, and $C_{n}$ are the time-dependent probability
amplitudes for finding the entire system in its corresponding states.

Inserting the wavefunction (\ref{4a1}) into the Schr\"{o}dinger
equation with the governing Hamiltonian in Eq.~(\ref{2a9}), we find
a system of coupled linear differential equations for the amplitudes
\begin{subequations}
\label{4a2}
\begin{eqnarray}
i\dot{U}_{k} &=&\Omega _{k}U_{k}+\sum_{jq}g_{q[j]}u_{k}^{\ast }\left(
j\right) B_{q[j]}+\sum_{n}u_{ek}^{\ast }\beta _{n}C_{n}\text{,} \\
i\dot{B}_{q[j]} &=&\omega _{q[j]}B_{q[j]}+\sum_{k}g_{q[j]}u_{k}\left(
j\right) U_{k}\text{,} \\
i\dot{C}_{n} &=&\nu _{n}C_{n}+\sum_{k}\beta _{n}u_{ek}U_{k}\text{,}
\end{eqnarray}%
where the overdot indicates the derivative with respect to time. The
evolution of $U_{k}$ is coupled to $B_{q[j]}$ and $C_{n}$ via the coupling
constant $g_{q[j]}u_{k}^{\ast }\left( j\right) $ and $u_{ek}^{\ast }\beta
_{n}$. To remove the high-frequency effect, we make the following
substitution
\end{subequations}
\begin{eqnarray}
U_{k}\left( t\right) &=&\phi _{k}\left( t\right) e^{-i\Omega _{k}t}\text{, }
\\
B_{q[j]}\left( t\right) &=&b_{q[j]}^{\prime }\left( t\right) e^{-i\omega
_{q[j]}t}\text{,}  \notag \\
C_{n}\left( t\right) &=&c_{n}\left( t\right) e^{-i\nu _{n}t}\text{.}  \notag
\end{eqnarray}%
Then in the interaction picture, equation~(\ref{4a2}) is rewritten as
\begin{eqnarray*}
i\dot{\phi}_{k} &=&\sum_{jq}g_{q[j]}u_{k}^{\ast }\left( j\right)
b_{q[j]}^{\prime }e^{-i\Delta _{qk}^{[j]}t}+\sum_{n}u_{ek}^{\ast }\beta
_{n}c_{n}e^{-i\delta _{nk}t}, \\
i\dot{b}_{q[j]}^{\prime } &=&g_{q[j]}\sum_{k}u_{k}\left( j\right) \phi
_{k}e^{i\Delta _{qk}^{[j]}t}, \\
i\dot{c}_{n} &=&\sum_{k}\beta _{n}u_{ek}\phi _{k}e^{i\delta _{nk}t},
\end{eqnarray*}%
where $\Delta _{qk}^{[j]}=\omega _{q[j]}-\Omega _{k}$, $\delta
_{nk}=\nu _{n}-\Omega _{k}$ are the detunings between the system S
and the environment B. Formally integrating the equations for
$b_{q[j]}$ and $c_{n}$ in the above equations yields
\begin{subequations}
\label{4a3}
\begin{eqnarray}
b_{q[j]}^{\prime } &=&-ig_{q[j]}\sum_{k}u_{k}\left( j\right) \int \phi
_{k}\left( \tau \right) e^{i\Delta _{qk}^{[j]}\tau }d\tau \\
c_{n} &=&-i\sum_{k}\beta _{n}u_{ek}\int \phi _{k}\left( \tau \right)
e^{i\delta _{nk}\tau }d\tau .
\end{eqnarray}%
Inserting Eqs.\,(\ref{4a3}) into the equation for $\phi _{k}$, we obtain
the exact integro-differential equation
\end{subequations}
\begin{eqnarray}
\dot{\phi}_{n} &=&-\sum_{kjq}g_{q[j]}^{2}u_{n}^{\ast }\left( j\right)
u_{k}\left( j\right) e^{-i\Delta _{qn}^{[j]}t}\int_{0}^{t}\phi _{k}\left(
\tau \right) e^{i\Delta _{qk}^{[j]}\tau }d\tau  \notag \\
&&-\sum_{kq}\beta _{q}^{2}u_{en}^{\ast }u_{ek}e^{-i\delta
_{qn}t}\int_{0}^{t}\phi _{k}\left( \tau \right) e^{i\delta _{qk}\tau }d\tau ,
\label{4a4}
\end{eqnarray}%
where the coupling constants $g_{q[j]}$ and $\beta _{q}$ are assumed to be
real.

Suppose that at time $t=0$ there is no interaction between the
system S and the bath B, and the S+B is in the state $\left\vert
\Omega _{k}00\right\rangle $, corresponding to a single excitation
in the system S. Mathematically, solving Eq.~(\ref{4a4}) is to solve
a initial value problem. It is well known that such an initial value
problem can be solved by Laplace transform. Denoting the Laplace
transform of $\phi _{n}\left( t\right) $ by $\bar{\phi}_{n}\left(
s\right) $ and taking into account of the initial conditions $\phi
_{k}\left( 0\right) =\delta _{nk}$, we have
\begin{equation}
\bar{\phi}_{n}=\frac{1}{s+I\left( s\right) },  \label{4a5}
\end{equation}%
where
\begin{equation*}
I\left( s\right) =\sum_{q}\left[ \sum_{j}\frac{g_{q[j]}^{2}\left\vert
u_{n}\left( j\right) \right\vert ^{2}}{s+i\Delta _{qn}^{[j]}}+\frac{\beta
_{q}^{2}\left\vert u_{en}\right\vert ^{2}}{s+i\delta _{qn}}\right] .
\end{equation*}%
The roots of the denominator in Eq.(\ref{4a5}) can be split into a
sum of the singular and principal value parts. The principal part is
merely included into redefinition of the energy. The singular part
gives the decay rate due to the coupling to the bath B. Under the
Wigner--Weisskopf approximation \cite{scully}, the system decay is
dominantly exponential with
rate%
\begin{equation}
\Gamma _{n}=\pi \sum_{j}\left\vert u_{n}\left( j\right) \right\vert
^{2}\Lambda_{j}\left( \Omega _{n}\right) +\pi \left\vert
u_{en}\right\vert ^{2}\Lambda_{A}\left( \Omega _{n}\right).
\label{4a6}
\end{equation}%
This rate is proportional to the modulus square of the coupling between S
and B.

\begin{figure}[tbp]
\includegraphics[width=8 cm]{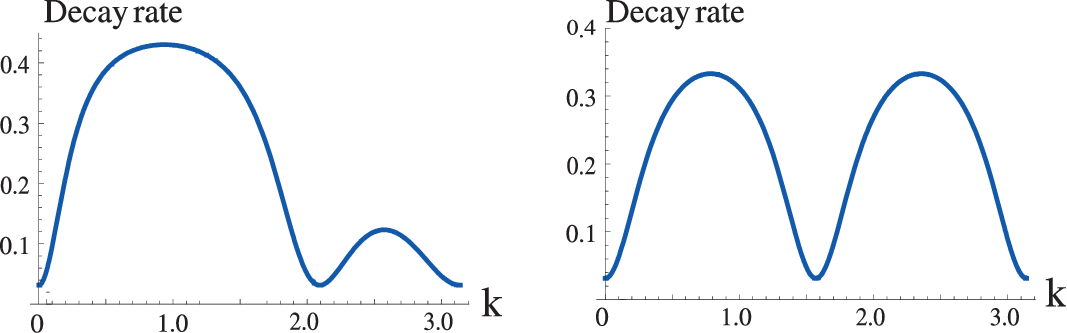}
\caption{(Color online) The decay rate via wavenumber $k$ starting
from one eigenstate of the system S, with parameters $g=0.1, \beta
=0.4,J=1.5$ , (a) $\omega _{C}=5,\Omega =6$, (b) $\omega _{C}=\Omega
=5$, which are in units of $\xi $.} \label{fig4:1}
\end{figure}

The functions in Eq.(\ref{4a6})
\begin{eqnarray*}
\Lambda_{j}\left( \omega \right) &=&\sum_{q}g_{q[j]}^{2}\delta
\left( \omega
-\omega _{q[j]}\right) \\
\Lambda_{A}\left( \omega \right) &=&\sum_{q}\beta _{q}^{2}\delta
\left( \omega -\nu _{q}\right)
\end{eqnarray*}%
are called the reservoir response (memory) functions, which are the
spectral densities of the states $b_{q[j]}^{\dag }\left\vert
G\right\rangle $ and $d_{q}^{\dag }\left\vert G\right\rangle $
weighted with the coupling strengths $g_{q[j]}^{2}$ and $\beta
_{q}^{2}$ respectively. The memory function $\Lambda\left( \omega
\right) $ characterizes the spectral shape of the reservoir. Since
the states in the reservoir are very dense (continuum), one can
replace the sums over $q$ by integrals, for instance,
$\sum_{q}\rightarrow 2V/(2\pi)^{3}\int q^{2}\sin\theta d\phi d\theta
dq$. Then the memory functions read
\begin{eqnarray*}
\Lambda_{j}\left( \omega \right) &\rightarrow &g_{[j]}^{2}\left(
\omega \right)
\rho _{\lbrack j]}\left( \omega \right) \text{,} \\
\Lambda_{A}\left( \omega \right) &\rightarrow &\beta ^{2}\left(
\omega \right) \rho _{A}\left( \omega \right) \text{,}
\end{eqnarray*}%
where $g_{[j]}^{2}\left( \omega \right) $ are the coupling constants
to states $b_{\omega \lbrack j]}^{\dag }\left\vert G\right\rangle $,
whose density of state is given by $\rho _{\lbrack j]}\left( \omega
\right) $. Obviously, the decay rate is determined by two factors,
1) the form of $\Lambda\left( \omega \right) $; 2) the width of the
reservoir, which depicts the overlap of the eigenfrequencies of the
system and the spectral shape of the reservoir. Usually, it is
assumed that the reservoir is spectrally flat and the frequencies of
the system of interest is deeply embedded in the continuum,
therefore, the argument of the spectral density is replaced by the
eigenfrequency of the system $\Omega _{n}$ in Eq. (\ref{4a6}). For
the sake of simplicity, we further assume that the coupling constant
$g_{q[j]}$ is independent of the energy $\omega _{q[j]}$ as well as
the index $j$, i.e. let $\Lambda_{[j]}(\omega)=g(\omega)$, and let
$\Lambda _{A}(\omega)=\beta _{A}(\omega)$. The constant coupling
gives us the explicit relationship between the decay rate and the
wavenumber of the system S
\begin{equation}
\Gamma _{n}=\pi g^{2}+\pi \left( \beta _{A}^{2}-g^{2}\right) \left\vert
u_{en}\right\vert ^{2},  \label{4a7}
\end{equation}%
where the normalization of the eigenstates $\left\vert \Omega
_{n}\right\rangle $ has been used in deriving above equation. Obviously,
when $\beta _{A}=g$, the decay rate is a constant for any eigenstates of the
system S. For a given wavenumber $n$, when $\beta _{A}$ is smaller than $g$,
the rate decreases, which opens up a possibility of prolonging the lifetime
of the single-excitation in the system S. In Fig.\ref{fig4:1}, we plot the
decay rate $\Gamma _{k}$ as a function of the wavenumber $k$.
It indicates that the widths of the eigestates in Eq.(\ref{3a4}) are
different, which is a periodical function of the wavenumber $k$, and
the detuning between the TLS and the resonator biases the symmetry
of the lineshape. By substituting the condition in Eq.(\ref{3a9})
into the expression of the probability amplitude $u_{e\kappa }$, the
decay rate of a bound state has the form $\Gamma _{\kappa }=\pi
g^{2}+\pi \left( \beta _{A}^{2}-g^{2}\right) \left[ 1-\left(
J^{2}+2\xi ^{2}\sinh 2\kappa \right) ^{-1}\right] $. It indicates
that the decay rate is an increasing function of $\kappa $, and when
the resonator resonates with the TLS, $\Gamma _{\kappa }=\pi \left(
2g^{2}+\beta _{A}^{2}\right) /3$. Since the value of the imaginary
wavenumber is determined for given parameters $\left\{ J,\xi ,\omega
_{C}-\Omega \right\} $, the decay rate of the bound state is a given
constant.

%%%%%%%%%%%%%%%%%%%%%%%%%%%%%%%%%%%%%%%%%%%%%%%%%%%%%%%%%%%%
\section{\label{Sec:5} dissipation on scattering amplitude}
%%%%%%%%%%%%%%%%%%%%%%%%%%%%%%%%%%%%%%%%%%%%%%%%%%%%%%%%%%%%

To see the impact of dissipation of the system on the
transmission/reflection coefficients, we begin with Hamiltonian
(\ref{2a9}) in the configuration space. Here, it is assumed that
each bath attached to its resonator is identical and the coupling
strength between the resonator and its corresponding bath is
independent of the location index $j$. The state at arbitrary time is
a superposition of four parts: the photon at the jth cavity with
atom in the ground state $a_{j}^{\dag }\left\vert G\right\rangle $,
no photon in all cavities with atom in the excited state $\sigma
_{+}\left\vert G\right\rangle $, the photon in the $q$th mode of the
bath attached to the $j$th resonator $b_{q[j]}^{\dag }\left\vert
G\right\rangle $ and that of the bath coupled to the TLS
$c_{q}^{\dag }\left\vert
G\right\rangle $%
\begin{eqnarray}
\left\vert \Psi \left( t\right) \right\rangle &=&\sum_{j}\Phi _{jk}\left(
t\right) a_{j}^{\dag }\left\vert G\right\rangle +\sum_{qj}A_{q[j]}\left(
t\right) b_{q[j]}^{\dag }\left\vert G\right\rangle  \label{5a1} \\
&&+\Phi _{ek}\left( t\right) \sigma _{+}\left\vert G\right\rangle
+\sum_{q}D_{q}\left( t\right) c_{q}^{\dag }\left\vert G\right\rangle .
\notag
\end{eqnarray}%
The Schr\"{o}dinger equation results in a system of coupled linear
differential equations for the amplitudes
\begin{subequations}
\label{5a2}
\begin{eqnarray}
i\dot{\Phi}_{jk} &=&\omega _{C}\Phi _{jk}-\xi \left( \Phi _{j+1k}+\Phi
_{j-1k}\right) +J\Phi _{ek}\delta _{0j}+\sum_{q}g_{q}A_{q[j]}, \\
i\dot{\Phi}_{ek} &=&\Omega \Phi _{ek}+J\Phi _{0k}+\sum_{q}\beta _{q}D_{q}
\notag \\
i\dot{A}_{q[j]} &=&\omega _{q}A_{q[j]}+g_{q}\Phi _{jk},  \notag \\
i\dot{D}_{q} &=&\beta _{q}\Phi _{ek}+\nu _{q}D_{q},  \notag
\end{eqnarray}%
where the dot denotes a derivative with respect to time. Applying
the Fourier transform and expressing $A_{q[j]}$, $D_{q}$ with $\Phi
_{jk}$, $\Phi _{ek}$ respectively, the equations of motion in the
frequency domain are obtained in the reduced dimensionality
\end{subequations}
\begin{equation}
\left( E-\omega _{C}-\delta E\right) \psi _{jk}=-\xi \left( \psi
_{j+1k}+\psi _{j-1k}\right) +\frac{J^{2}\psi _{0k}\delta _{0j}}{E-\Omega
-\Delta E},  \label{5a3}
\end{equation}%
where $E$ is the eigenenergy of the single-photon wave in the whole
system. The eigenfrequency of the $j$th resonator of the ideal CRW
is renormalized into $\omega _{C}+\delta E$, and the transition
energy of the TLS is also renormalized into $\Omega +\Delta E$
\begin{equation}
\delta E=\sum_{q}\frac{g_{q}^{2}}{E-\omega _{q}}, \quad
\Delta E=\sum_{q}\frac{\beta _{q}^{2}}{E-\nu _{q}},  \label{5a4}
\end{equation}%
which is the influence of the bathes on the state of each resonator
and the TLS. The singular part of $\delta E$ and $\Delta E$ yields
the dissipation factors $\gamma _{c}=\pi g^{2}\left( \omega
_{C}\right) \rho _{c}\left( \omega _{C}\right) $ and $\gamma
_{A}=\pi \beta \left( \Omega \right) \rho _{A}\left( \Omega \right)
$. The real part of $\delta E$ and $\Delta E$ contributes to the
Lamb shift of the levels and change of $\Omega$, respectively, which
is merely included in the redefinition of the energy. Then Eq.(\ref{5a3})
is reduced to the following set of equations
\begin{equation}
\left( E-\omega _{c}+i\gamma _{c}\right) \psi _{jk}=-\xi \left( \psi
_{j+1k}+\psi _{j-1k}\right) +\frac{J^{2}\psi _{0k}\delta _{0j}}{E-\Omega
+i\gamma _{A}}  \label{5a5}
\end{equation}

We begin our analysis from the case with coupling strength $J=0$. In
this case, Eq.(\ref{5a3}) has a complex dispersion curve that can be
identified with the plane-wave solution $\Phi _{jk}\propto
e^{i\left( kj-Et\right) }$, where $E=E_{r}+iE_{i}$ is the complex
energy and $k$ is the wave momentum inside the CRW. The real and
imaginary parts of energy $E$ satisfy $E_{r}=\Omega _{k}$,
$E_{i}=\gamma _{c}$. When only one resonator is initially excited
$\left\vert \Psi \left( 0\right) \right\rangle =\left\vert
n\right\rangle $ (say $n=0$), where $\left\vert n\right\rangle $ is
the Wannier state localized at the site $n$, the field profile at
time $t$ is
given by%
\begin{equation}
\left\vert \Psi \left( t\right) \right\rangle =e^{-i\omega _{c}t-\gamma
_{c}t}\sum_{kq}J_{q}\left( 2\xi t\right) e^{iq\pi /2}e^{-ikq}\left\vert
k\right\rangle  \label{5a6}
\end{equation}%
where $\left\vert k\right\rangle $ is the Bloch state. The amplitude at the
site $l$ reads
\begin{equation}
\Phi _{lk}=e^{-i\omega _{c}t-\gamma _{c}t}J_{l}\left( 2\xi t\right) e^{il\pi
/2}  \label{5a7}
\end{equation}%
where $J_{l}\left( x\right) $ is a Bessel function of the first kind
of integer order $l$. Obviously, leakage rate in each resonator
$\gamma _{c}$ influences the free propagation of the single photon.
The distance that the photon travels along the 1D resonator
waveguide is depicted by the product of the group velocity and
$\gamma _{c}^{-1}$.

\begin{figure}[tbp]
\includegraphics[width=8 cm]{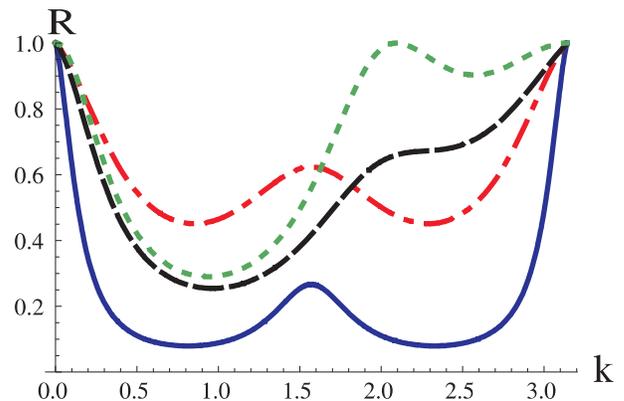}
\caption{(Color online) The reflection coefficient versus the
wavenumber $k$ . The parameters are in units of $\protect\xi $ and
are set as follows: $ \protect\gamma _{c}=0.1,\protect\gamma
_{A}=0.4,J=0.8,\ \omega _{c}=\Omega =5$ for blue solid line; $
\gamma _{c}=0.1, \gamma _{A}=0.4,J=1.5, \omega _{c}=\Omega =5$ for
red dot-dashed line; $\gamma _{c}=0.1, \gamma _{A}=0.4,J=1.5, \omega
_{c}=5,\Omega =6$ for black dashed line; $\gamma _{c}=0.1, \gamma
_{A}=0.1,J=1.5, \omega _{c}=5,\Omega =6$ for green dotted line. }
\label{fig5:1}
\end{figure}

We now consider the case with coupling strength $J\neq 0$. Due to
the coupling between the atom and the $0$th resonator, a complex
$\delta $-like potential stands in the way that single photon
travels. Consequently, the photon experiences scattering. We assume
that the single-photon has momentum $k$ initially. Within the
allowed distance of photon traveling along the 1D CRW, the
transmission and reflection amplitudes, $t$ and $r$, can be defined
via the asymptotes of the wave function
\begin{equation}
\psi _{jk}=\left\{
\begin{array}{c}
e^{ikj}+r_{k}e^{-ikj}\text{, }j<0 \\
s_{k}e^{ikj}\text{, \ \ \ \ \ \ \ }j>0.
\end{array}%
\right.   \label{5a8}
\end{equation}%
By substituting the asymptotes of the wave function into Equation
(\ref{5a5}) for $j=\pm 1,0$ sites, the reflection amplitude is
obtained as
\begin{equation}
r_{k}=\frac{J^{2}}{2i\xi \sin k\left( \Omega _{k}-\Omega \right)
-\left( \gamma _{A}-\gamma _{c}\right) 2\xi \sin k-J^{2}}.
\label{5a9}
\end{equation}%
Equation~(\ref{5a9}) shows that as long as the rate of decay to each
resonator $\gamma _{c}$ is equal to the decay rate of the TLS
$\gamma _{A}$, a resonant scattering occurs when the incident energy
of the single-photon is equal to $\Omega$, the transition energy of
the TLS, i.e. when the single-photon incident from the left
encounters the TLS, it is completely reflected back to the left, as
shown via the green dotted line in Fig.~\ref{fig5:1}.
For unequal $\gamma _{c}$ and $\gamma _{A}$, the decay rates lower
the peak of the resonance, and the width of the line shape is
broadened from $J^{2}/\left( 2\xi \sin k\right) $ for ideal system
$S$ to $\left[ \left( \gamma _{A}-\gamma _{c}\right) 2\xi \sin
k+J^{2}\right] /\left( 2\xi \sin k\right) $. In Fig.\ref{fig5:1}, we
plot the reflection coefficient as a function of the wavenumber $k$.
Comparing the blue solid line and the red dash-dot line with the
black dashed line and the green dotted line, it can be found that
the symmetry of the lineshape is determined by the transition energy
$\Omega $ of the TLS. As the coupling strength $J$ between the $0$th
resonator and the TLS goes to infinity, the reflection coefficient
approaches to one.

\section{\label{Sec:6}Conclusion}

In summary, we have studied the dissipative process of the CRW with
a TLS inside, originating from the coupling to the environment. Our
discussions are based on a simple decoherence scenario in which each
resonator and the TLS individually interact with its own
environment. Each environment is modeled as a continuum of harmonic
oscillators and  assumed to be on the vacuum state initially. Since
the coupling with the environment is generally weak compared to the
system of interest, the Wigner--Weisskopf approximation applies, and
the lifetime of an excitation in the system S is thus obtained,
which represents the time-scale of the transition from quantum to
classical behavior. We further investigate the impact of the
dissipation on the transport property of single-photon along the 1D
CRW for identical bathes attached to the CRW. The dissipation lowers
the peak of the resonance and broadens the width of the line shape
of the reflection spectrum except the case that magnitude of the
leakage rate is equal to the decay rate of the TLS. The leakage rate
in each resonator not only influences the free propagation of the
single photon, but also results in the inelastic scattering of the
single-photon together with the dissipation of the TLS.

This work is supported by New Century Excellent Talents in University
(NCET-08-0682), NSFC No.~10775048, and No.~10704023, NFRPC 2007CB925204, and
Scientific Research Fund of Hunan Provincial Education Department No.~09C638
and No.~09B063.

\end{document}